\documentclass[reprint,twocolumn,superscriptaddress,showpacs,nofootinbib,notitlepage,preprintnumbers]{revtex4-2}

\usepackage{graphicx}
\usepackage{hyperref}
\usepackage[dvipsnames]{xcolor}
\usepackage{amsmath,amsthm,amssymb}
\usepackage{comment}
\usepackage{enumerate}
\newcommand{\todo}{\colorbox{pink}{\textsc{Todo}}}

\newcommand{\yoshio}{\colorbox{Yellow}{\textsc{Yoshio}}}

\renewcommand{\Re}{\mathrm{Re}\,}
\renewcommand{\Im}{\mathrm{Im}\,}
 
\begin{document}
\preprint{LA-UR-23-31682}
\preprint{INT-PUB-24-002}

\title{Contour deformations for non-holomorphic actions}

\author{Scott Lawrence}
\email{srlawrence@lanl.gov}
\affiliation{Los Alamos National Laboratory Theoretical Division T-2, Los Alamos, NM 87545, USA}
\author{Semeon Valgushev}
\email{semeonv@iastate.edu}
\affiliation{Department of Physics and Astronomy, Iowa State University, Ames, IA, 50011, USA}
\author{Jianan Xiao}
\email{yoshi2020@keio.jp}
\affiliation{Department of Physics, Keio University, Yokohama, 223-8522, Japan}
\author{Yukari Yamauchi}
\email{yyama122@uw.edu}
\affiliation{Institute for Nuclear Theory, University of Washington, Seattle, WA 98195, USA}

\date{\today}

\begin{abstract}
We show how contour deformations may be used to control the sign problem of lattice Monte Carlo calculations with non-holomorphic actions. Such actions arise naturally in quantum mechanical scattering problems. The approach is demonstrated in conjunction with the holomorphic gradient flow. As our central example we compute the real-time evolution of a particle in a one-dimensional analog of the Yukawa potential.
\end{abstract}

\maketitle

\section{Introduction}

Sign problems prevent lattice Monte Carlo calculations---which are otherwise a unique window into nonperturbative quantum systems---from accessing a variety of regimes. Most famous are the fermion sign problems that afflict the Hubbard model away from half-filling and QCD at finite density. Real-time calculations are similarly obstructed. Mitigating or evading these sign problems remains an active area of research~\cite{Aarts:2008rr,Langfeld:2016mct,Alexandru:2005ix,deForcrand:2006ec,Fodor:2001au,Allton:2002zi,Chandrasekharan:2013rpa,deForcrand:2006pv,Alexandru:2020wrj,Lawrence:2022afv}.

Expectation values in a theory with a sign problem are generally evaluated by reweighting. The desired physical expectation value is computed as a ratio of two expectation values with respect to a quenched Boltzmann factor:
\begin{equation}\label{eq:reweighting}
    \langle \mathcal O \rangle \equiv \frac{\int\!e^{-S} \mathcal O }{\int \!e^{-S}} \!=\!
    \frac{\int\! \mathcal O e^{-S}\big/\int\! e^{-\Re S}}{\int\! e^{-S}\big/\int\! e^{-\Re S}}
    \equiv
    \frac{\langle \mathcal O e^{-i \Im S}\rangle_Q}{\langle e^{-i \Im S} \rangle_Q}
    \text.
\end{equation}
Above, the action $S$ is complex (creating the sign problem), $\mathcal O$ is the desired observable, and the integration measure over all field configurations has been omitted for brevity. Quenched expectation values are denoted $\langle \cdot \rangle_Q$, and are defined as expectation values over the probability distribution $e^{-\Re S}$.

In this approach, the sign problem appears most prominently as a signal-to-noise problem in the computation of the denominator in Eq.~(\ref{eq:reweighting}). This denominator is termed the \emph{average phase}, denoted $\langle \sigma\rangle$, and used as a convenient measure of the difficulty of the sign problem. The origin of the signal-to-noise problem can be understood as follows: each sample is a complex number of unit modulus, but the expectation value itself is typically exponentially small in the spacetime volume of the system being simulated. As a result, an exponentially large number of samples are required to resolve the denominator from zero, and therefore to determine the sign of the physical expectation value $\langle \mathcal O \rangle$.

The method of contour deformations (originated in~\cite{Cristoforetti:2012su}; see~\cite{Alexandru:2020wrj} for a review) stems from the observation that the domain of integration in Eq.~(\ref{eq:reweighting}) consists of all real-valued field configurations, but particularly in lattice field theory, the observables and actions of interest are often well-behaved complex-analytic functions. We may therefore deform the contour of integration into the space of complex-valued field configurations, with Cauchy's integral theorem guaranteeing that integrals of holomorphic functions are not modified. Since physical expectation values are all defined by the partition function $\int e^{-S}$, and the action itself is holomorphic, we are able to perform such deformations without introducing any systematic bias. However, the average phase $\langle\sigma\rangle$ depends on the non-holomorphic integral $\int e^{-\Re S}$, and therefore the sign problem might be made better by such a procedure. In practice it has been found that the sign problem is often improved by orders of magnitude, and effort has focused on methods for obtaining good integration contours, particularly Lefschetz thimbles~\cite{Alexandru:2017czx,Cristoforetti:2012su,Alexandru:2015xva} and machine learning approaches~\cite{Lawrence:2022afv,Lawrence:2021izu,Alexandru:2018ddf,Alexandru:2018fqp,Alexandru:2018ngw,Mori:2017pne,Kashiwa:2020brj,Kashiwa:2023dfx}.

The contour deformation method depends on the analyticity of the action $S$ (and all observables of interest, or at least the product $e^{-S} \mathcal O$~\cite{Alexandru:2018ngw}). Without analyticity, Cauchy's integral theorem does not apply and the process of deforming the integration contour will generically change expectation values. This raises the question of what can be done to suppress sign problems that arise in nuclear systems, where the nucleon-nucleon potential is typically nonanalytic. Recent work evades this issue by performing a high-degree approximate fit of the nuclear potential to analytic functions~\cite{Kanwar:2023otc}, thus providing a holomorphic action which is compatible with contour deformation methods.

In this paper we suggest another approach for handling nonanalytic actions. Treating the action in question as only a function of real field configurations, we perform explicit analytic continuation to a suitable Riemann surface. Thus we define a holomorphic action on some complex manifold, typically not homeomorphic to $\mathbb C$. Contour deformations can now be performed within this manifold, with all physical expectation values protected by Cauchy's integral theorem as usual.

This paper is structured as follows. Section~\ref{sec:sheets} shows how via appropriate analytic continuation, various actions may be considered as analytic functions on complex manifolds with multiple sheets. The rules for contour deformation are derived in the same section. The holomorphic gradient flow, a commonly used tool for finding good contours for integration, is generalized in Section~\ref{sec:hgf} to the case of multiple sheets. In Section~\ref{sec:yukawa} we apply these techniques to quantum-mechanical scattering in one dimension. We conclude in Section~\ref{sec:discussion} with a discussion of some open questions.

\section{Contour deformations on multiple sheets}\label{sec:sheets}

The perspective taken in this work is that, when presented with a non-holomorphic action $S_{\mathrm{orig}}(z,\bar z)$, we should consider only its restriction $S_{\mathbb R}(x)$ to real-valued field configurations $x$. Doing so loses no physical information: the partition function itself is typically written as an integral only over such configurations. At the same time, after this restriction, we are now free to choose the behavior of a new function $S(z)$ to agree with $S_{\mathbb{R}}$ on real-valued configurations, but be holomorphic elsewhere. The manifold on which this new action is holomorphic may not be homeomorphic to $\mathbb C$, and we expect in general to encounter multiple branches.

To make this clear, let us begin with an unfairly simple example. In one complex dimension, consider the integral
\begin{equation}
    Z_{\mathrm{Gauss}}
    =
    \int_{-\infty}^{\infty} dz\,e^{-\alpha |z|^2}\text.
\end{equation}
This corresponds to the action $S_{\mathrm{orig}}(z,\bar z) = \alpha \bar z z$. The integrand is manifestly non-holomorphic, and we cannot invoke Cauchy's integral theorem. As discussed, this function is defined on $\mathbb C$, but we will only consider its restriction $S_{\mathbb R}(x)$ to the real line $\mathbb R$. All other information contained in $S_\mathrm{orig}$ is physically irrelevant.

We may now select any function $S(z)$ that agrees with $S_{\mathbb R}(x)$ at real values $z = x$. The reader will not be surprised by the choice
\begin{equation}
    S(z) = \alpha z^2\text.
\end{equation}
Note that although we may write the restricted action as $S_{\mathbb{R}}(x)=\alpha x^2$, these two actions are not the same object. The restricted action has domain $\mathbb R$, and the new action, which is constructed for computational convenience, has domain $\mathbb C$. The distinction will become more important as we proceed to less trivial systems.

Using the new action, we now write our integral as
\begin{equation}
    Z_{\mathrm{Gauss}}
    =
    \int_{-\infty}^{\infty} dz\,e^{-\alpha  z^2}\text,
\end{equation}
and it is easy to find a contour deformation that exactly removes any sign problem introduced by the choice of complex $\alpha$.

Let us now repeat these steps, as closely as possible, for a less trivial integral. Consider
\begin{equation}
    Z_{\mathrm{abs}} = \int_{-\infty}^{\infty} e^{-\alpha |z|}\text.
\end{equation}
The integrand is now not just non-holomorphic, but has a singularity (in this case a cusp) on the real line itself, indicating that no unique analytic continuation to $\mathbb C$ is possible. We will see that evading this requires the introduction of two sheets.

Before proceeding, it's worth noticing a simple proof-of-principle argument that contour deformation methods can be productively applied in this case. We might split the integral up into two terms, like so:
\begin{equation}
    Z_{\mathrm{abs}} =
    \int_{-\infty}^{0}
    e^{\alpha z}
    +
    \int_{0}^{\infty} e^{-\alpha z}\text.
\end{equation}
In performing this split, we were able to eliminate any reference to $\bar z$. We now have a sum of two integrals, and we may deform those contours (subject to the requirement that each must terminate at the origin) to remove the sign problem. However, in the case of a many-dimensional integral, this viewpoint is unworkable: we will have exponentially many such terms in the number of degrees of freedom.

\begin{figure}
    \centering\includegraphics[width=0.9\linewidth]{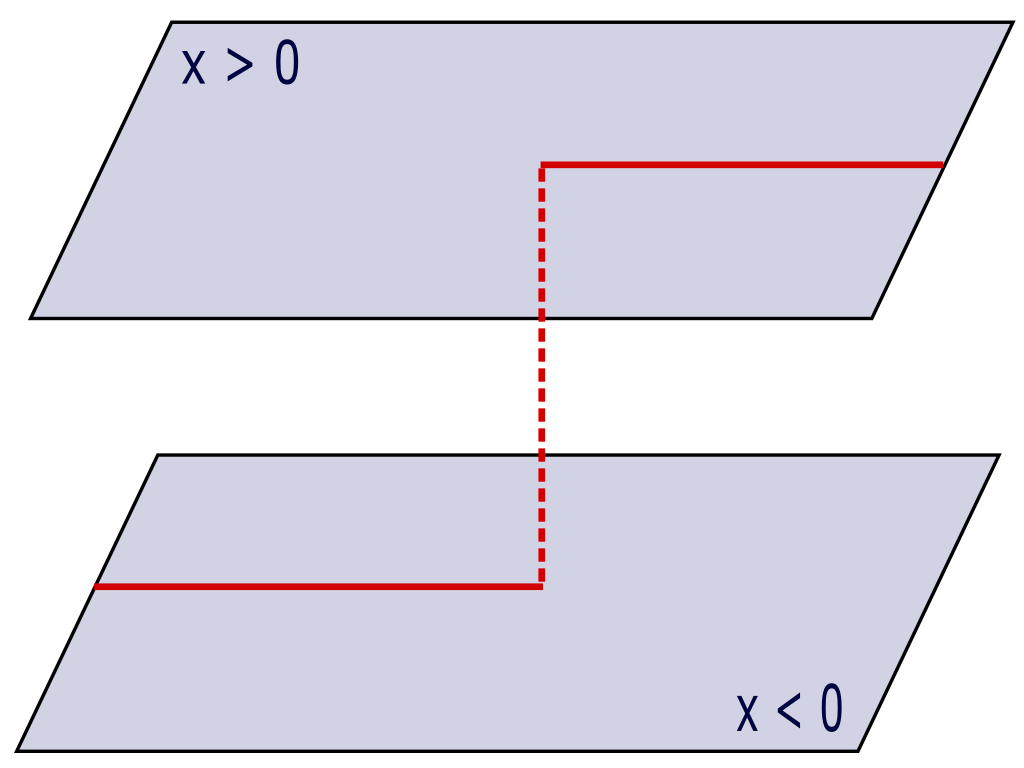}
    \caption{Cartoon of the two sheets for actions depending only on $|z|$. Note that the ``real line''---that is, the original integration contour---begins on the bottom sheet and ends on the top. As the origin is the only point connecting the two, any contour deformation must go through the origin.\label{fig:cartoon}}
\end{figure}

We return instead to the original plan. The original action, when restricted to the real line, yields
\begin{equation}
    S_{\mathbb{R}}(x)
    =
    \begin{cases}
        \alpha x& x \ge 0\\
        - \alpha x& x < 0
    \end{cases}
    \text.
\end{equation}

We cannot hope to find an analytic function $S(z)$ that agrees everywhere on the real line. However, we can perform analytic continuation\footnote{Constructively, analytic continuation is performed by beginning at a point, selecting a path, and evolving the Cauchy-Riemann equations along that path.} beginning from a chosen point $x_0 \ne 0$. In a finite neighborhood around any point other than the origin, $S_{\mathbb R}(x)$ is clearly analytic, and in fact possesses a unique analytic continuation to the remainder of the complex plane---\emph{including}, in this case, the origin.

We therefore see that we have two analytically continued actions:
\begin{equation}
    S_+(z) = \alpha z \text{ and } S_-(z) = -\alpha z
    \text.
\end{equation}
If we wish to view these as one function, we must add a single discrete degree of freedom to indicate whether we are on the sheet defined by analytic continuation from positive real $z$, or that defined by continuation from negative real $z$. The result is shown in Figure~\ref{fig:cartoon}. The domain of $S(z,\mathrm{sheet})$ is two copies of the complex plane $\mathbb C$, identified at the origin.

An alternate perspective may be enlightening. For an analytic function $f(\cdot)$, the function $f[\sqrt{(x-a)(x+a)]})$ has two branches, connected by the line segment $x \in (-a,a)$. In the limit as $a \rightarrow 0$, the connection between the two branches contracts to a point, while the branches remain distinct. This is precisely the structure described above and portrayed in Figure~\ref{fig:cartoon}.

The original contour of integration is shown in Figure~\ref{fig:cartoon} in red. It begins on the `negative' sheet (the analytic continuation from $x < 0$), goes to the origin, and then continues on the `positive' sheet. As usual, we may perform any continuous contour deformation without changing the integral. In this case, the requirement of continuity imposes a particular restriction: because the contour must begin on one sheet and end on the other, it must always go through the origin, which is the only point connecting the two sheets.

This framework is agnostic to the sort of singularity being introduced, to the location of the singularity, and to the number of singularities. 

\section{Holomorphic gradient flow}\label{sec:hgf}

\begin{figure*}
\centering
\includegraphics[width=0.48\linewidth]{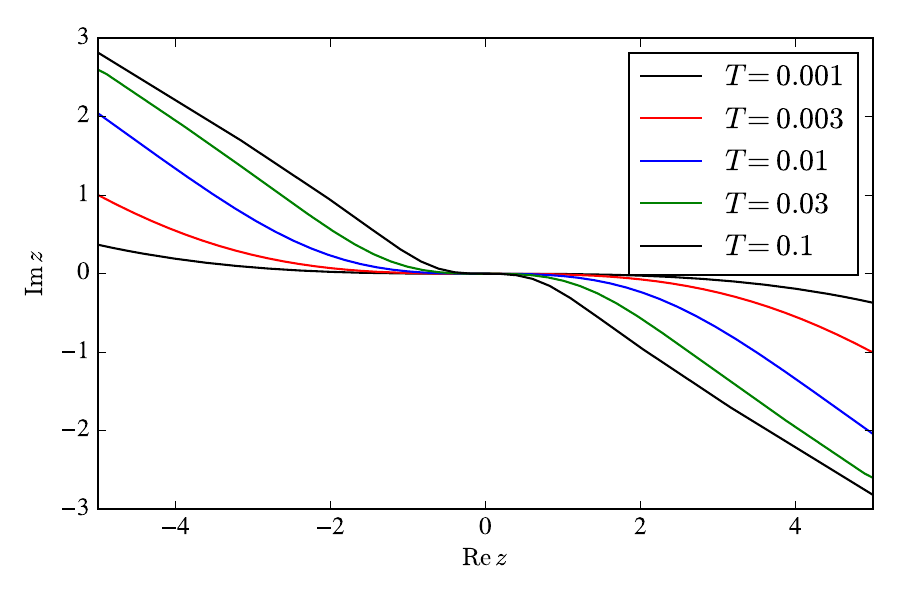}\hfill
\includegraphics[width=0.48\linewidth]{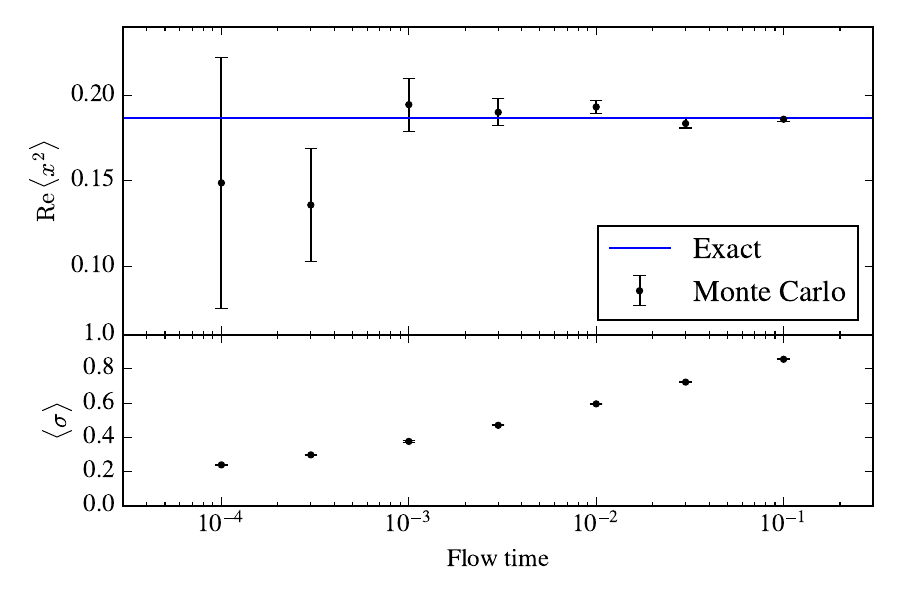}
\caption{Treating the sign problem of Eq.~(\ref{eq:toy-cubic}) via modified holomorphic gradient flow. At left, flowed integration contours for $T=0.001$ through $T=0.1$. The distinction between the two branches is not shown explicitly. At right, the average sign and expectation value of $z^2$ as a function of flow time. Each data point corresponds to $10^4$ samples.\label{fig:cubic}}
\end{figure*}

To select an integration contour with an acceptable average phase, we turn to the strategy of using the holomorphic gradient flow as described in~\cite{Alexandru:2015sua}. Each point on the initial integration contour (in most cases, including in this work, the real line) is evolved under the differential equation
\begin{equation}\label{eq:hgf}
    \frac{dz}{dt} = \overline{\frac{\partial S}{\partial z}}\text.
\end{equation}
The set of solutions at some fixed time $T$ (termed the flow time) constitutes a new integration contour. Empirically, such contours have been found to have dramatically improved average phase over the real line in many circumstances. Partial explanation may be found in the fact that the real part of the action increases along lines of flow:
\begin{equation}
    \frac{d S}{d t} = \left|\frac{\partial S}{\partial z}\right|^2\text.
\end{equation}
Thus, at least in the limit of $\hbar \rightarrow 0$ where the Jacobian $J$ in the effective action $S_{\mathrm{eff}}=S-\hbar\log\det J$ may be ignored, this flow equation directly suppresses the quenched partition function.

In the usual context, the differential equation (\ref{eq:hgf}) is viewed as defining a function $z(t)$ taking values in the complex plane $\mathbb C^N$. When the action has been written as a holomorphic function of a Riemann surface other than the complex plane, $z(t)$ now takes values on that surface. The right-hand side of (\ref{eq:hgf}) is still valued in $\mathbb C^N$, as that is the tangent space of the Riemann surface.

As a demonstration of this approach, Figure~\ref{fig:cubic} shows several computations of the expectation value $\langle x^2 \rangle$ for different flow times, with respect to the action
\begin{equation}\label{eq:toy-cubic}
S = \alpha |z|^3
\text,
\end{equation}
with $\alpha = i$ taken to maximize the difficulty of the sign problem.
Note that the expectation value in question does not converge on the real plane. It is reasonable to define it by analytic continuation in the parameter $\alpha$; as discussed in~\cite{Lawrence:2022afv}, contour deformations access exactly this analytic continuation. With such an interpretation, we have
\begin{equation}
\langle x^2 \rangle =
    \frac{\int_{-\infty}^\infty e^{-i |z|^3} z^2 dz}{\int_{-\infty}^\infty e^{-i |z|^3} dz}
    = \left[
    \frac{-3i \pi}{\Gamma(-\frac 1 3)}
    + \frac{\Gamma(\frac 1 3)}{2}
    \right]^{-1}
    \text.
\end{equation}
At vanishing flow time, the average phase is exactly $0$ for this action; even flow times as short as $T = 10^{-4}$ are seen to produce a manageable average phase. The average phase is roughly linear in the logarithm of the flow time, up to $T=10^{-1}$.

\section{Yukawa scattering}\label{sec:yukawa}

For a nontrivial demonstration, we turn to a one-dimensional scattering problem. The Yukawa potential in three dimensions is derived by integrating out a massive boson mediating the interaction. In one dimension the same procedure yields
\begin{equation}
    \label{eq:yukawa}
    V(x) = - g^2 |x| e^{-m |x|}
    \text.
\end{equation}
We consider this to be the one-dimensional analog of the Yukawa potential, written in terms of the center-of-mass coordinate $x$.

Because of the low dimensionality, scattering from this potential is readily computable by a variety of numerical methods.  In this work we adopted the approach where the time-evolution operator $e^{-i H t}$ is expanded via second-order Trotter-Suzuki as a product of time-evolutions under the free (kinetic) Hamiltonian and the potential term:
\begin{equation}\label{eq:trotter}
    e^{-i H t} \approx
    \left(e^{-i V(x) \frac{\delta}{2}} e^{-i \frac{p^2}{2M} \delta} e^{-i V(x) \frac{\delta}{2}}
    \right)^{t/\delta}
    \text.
\end{equation}
Discretizing a wavefunction on a one-dimensional lattice, the evolution under the potential term is easily performed by multiplication:
\begin{equation}\Psi_{t+\delta}(x) = e^{-i V(x) \delta} \Psi_{t}(x)
\text.
\end{equation}
Evolution under the kinetic term is accomplished by first performing a fast Fourier transform to obtain the wavefunction in momentum space.
The required number of lattice sites, and hence the resource consumption of the algorithm, is exponential only in the number of spatial dimensions (here $1$).

With a repulsive potential strength (in units where the mass of the particle is $M=1$) of $g^2 = -150$, a range determined by $m = 10$, and an initial wavefunction defined by the Gaussian wavepacket
\begin{equation}
\Psi_{i}(x)
\propto e^{-\frac{(x - x_0)^2}{2 \sigma^2} - i p_0 x}
\text,
\end{equation}
with $x_0 = -1$, $\sigma = 0.2$, and $p_0 = 1$,
the transmission coefficient is found by this method to be slightly smaller than $0.5$. Throughout this section we will retain these same parameters and initial state.

\begin{figure*}
\centering
\includegraphics[width=0.48\linewidth]{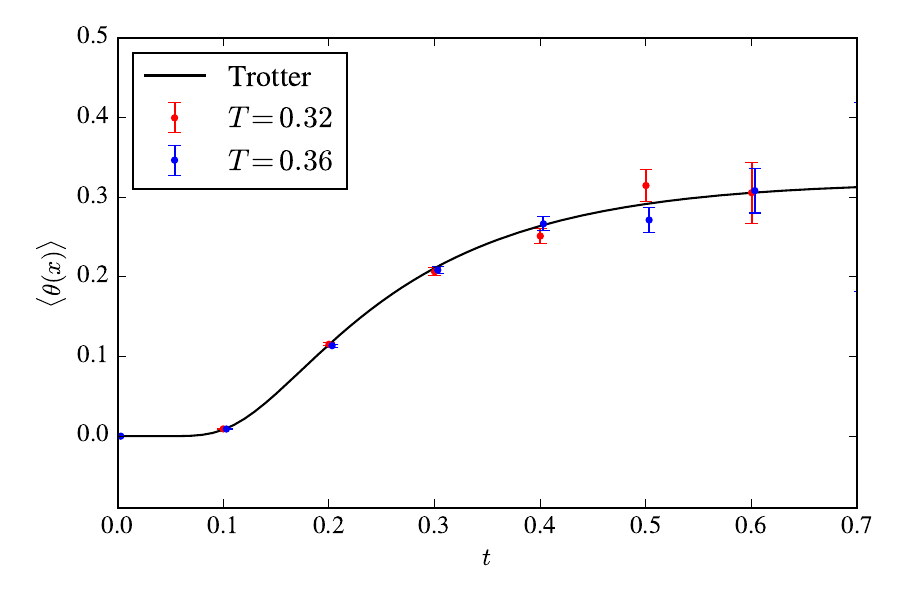}\hfill
\includegraphics[width=0.48\linewidth]{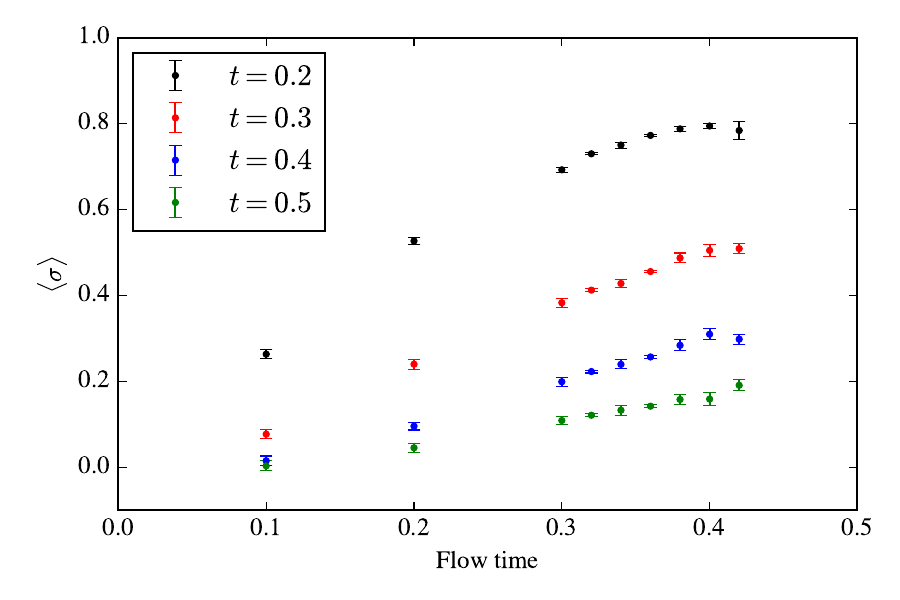}
\caption{Computation of the Yukawa transmission amplitude via holomorphic gradient flow. At left, the probability of the particle having passed the center of the potential is shown as a function of flow time, computed with two different holomorphic gradient flow times and with exact time-evolution. At right, the average phase for lattices with different time extents, as a function of flow time. Each data point corresponds to $10^4$ samples except for $T=0.32, 0.36$ which correspond to $10^5$ samples.\label{fig:yukawa}}
\end{figure*}

The primary purpose of this section is to show how the same calculation can be performed via lattice Monte Carlo methods, suitably augmented. We want to evaluate time-dependent expectation values in a fixed initial state:
\begin{equation}\label{eq:goal}
    F(t) \equiv \langle \Psi_i | e^{i H t} \mathcal O e^{-i H t}| \Psi_i \rangle
    \text.
\end{equation}
In scattering problems, the initial state will typically be a Gaussian wavepacket as given above; however the formalism is general. 
The Hamiltonian of our system is that of an otherwise free particle encountering a potential well centered around $x=0$:
\begin{equation}
    H = \frac{p^2}{2M} - g^2 |x| e^{-m|x|}\text.
\end{equation}
We may easily evaluate matrix elements of the time evolution operator in the limit of small times $\delta$ via the Trotter-Suzuki expansion as in Eq.~(\ref{eq:trotter}). As a result, beginning from (\ref{eq:goal}), expanding the time-evolution operators as $e^{-i H t} = (e^{-i H \delta})^{t/\delta}$, and inserting the position-space resolution of the identity 
\begin{equation}
    I \propto \int dx \, | x \rangle \langle x |
\end{equation}
between every pair of operators, we arrive at a discrete path-integral expression for the time-dependence $F(t)$:
\begin{equation}\label{eq:pi}
    F(t) = \frac{\int \left[\prod_{n=0}^{2N} d x_n\right]e^{-S(x)} \mathcal O(x_N)}{\int \left[\prod_{n=0}^{2N} d x_n\right]e^{-S(x)}}
    \text,
\end{equation}
where the lattice action is given by
\begin{widetext}
\begin{equation}\label{eq:action}
    S(x) = -\log \Psi_i(x_0) - \log \Psi_i^\dagger(x_{2N}) 
    - i\sum_{n=0}^{N-1} \left[\frac{(x_n - x_{n+1})^2}{\delta} + \delta_n V(x_n)\right]
    + i\sum_{n=N+1}^{2N} \left[\frac{(x_{n-1} - x_{n})^2}{\delta} + \delta_n V(x_n)\right]
    \text.
\end{equation}
\end{widetext}
Note that
\begin{equation}
    \delta_n
    =
    \begin{cases}
        \delta/2 &n=0,2N\\
        \delta &\mathrm{otherwise}
    \end{cases}
\end{equation}
as we use the second-order Trotter-Suzuki expansion. 

To make Monte Carlo methods applicable we must rewrite this as an expectation value over a probability distribution. This is done by reweighting: the desired physical expectation value may be written as a ratio of two quenched expectation values:
\begin{equation}
\langle \mathcal O \rangle = \frac{\langle \mathcal O e^{-i \Im S}\rangle_Q}{\langle e^{-i \Im S}\rangle_Q}
    \text.
\end{equation}
As with the previous example, the denominator of this expression is zero: in order for the integrals in this expression to converge, the oscillations must be regulated by an appropriately chosen infinitesimal contour deformation away from the real plane.

This approach is only a slight modification of the lattice Schwinger-Keldysh formalism~\cite{Alexandru:2016gsd,Alexandru:2017lqr}. In that context, the initial state is defined by a density matrix proportional to $e^{-\beta H}$, which is represented by addition terms in the action defining the Euclidean evolution. In our case we begin with a pure state whose wavefunction is known analytically, and the Euclidean evolution is not needed.

We must now determine the operator $\mathcal O$ to be measured. We wish to determine the transmission coefficient; assuming that the incident wavepacket begins at $x < 0$, the most directly relevant observable is the probability of observing the particle at some $x > 0$. This probability is given by the operator $\hat T \equiv \theta(\hat x)$, equivalently defined by its matrix elements
\begin{equation}
    \langle x' | \hat T | x \rangle = \delta(x-x') \theta(x)\text.
\end{equation}
In the path-integral formalism the corresponding expectation value is just that of
\begin{equation}
    T(x) = \theta(x_t)\text.
\end{equation}
This is of course non-analytic, and as previously discussed would not be an acceptable observable for use with contour deformation methods. Happily, the branch structure of the Hadamard function is no different\footnote{This can be seen most clearly from $\theta(x) = \frac 1 2 \left(1 + \frac{x}{|x|}\right)$.} from that of $|x|$. The function $T(x)$, analytically continued to the complex manifold on which the action is defined, is simply an indicator function indicating which branch of $|x|$ is being sampled. Thus the probability of the particle being found at $x>0$ is precisely the probability of sampling the positive branch.

The discussion so far suffices to apply general contour deformations to the scattering problem. The holomorphic gradient flow itself is not directly applicable, however. To see this, consider the behavior of the flow when (at least) one $x_n$ is near $0$. From the potential term in the action, we find
\begin{equation}
    \frac{d x_n}{d T} = g^2 \theta(x_n) + \cdots
    \text,
\end{equation}
where terms subleading in $x_n$ have been dropped. It is apparent that the flow is discontinuous at $x_n = 0$ (except for the case $n=N$), and therefore that after any finite flow time, the integration contour is no longer continuous. The conditions of Cauchy's integral theorem not being satisfied, expectation values will no longer be independent of flow time.

We can remove this discontinuity---rendering the holomorphic gradient flow usable---by modifying the defining flow equation such that the velocity field falls to zero even at singular points of $\frac{\partial S}{\partial z}$. To this end we modify Eq.~(\ref{eq:hgf}) to
\begin{equation}\label{eq:hgf-regulated}
    \frac{dz}{dt} = \overline{\frac{\partial S}{\partial z}}
    s(x)
    \text,
\end{equation}
where $s(x)$ is a scale function which vanishes sufficiently quickly at singular points to ensure the continuity of the contour. This function is far from unique---what constraints on $s(x)$ exist follow from the usual requirement that the deformation of the contour be a continuous process. For the scattering problem at hand we find that the following choice works sufficiently well:
\begin{equation}\label{eq:hgf-flow-scale}
    s(x) =    e^{-\frac{|x|}{\sqrt{m}}} I_1\left(\frac{|x|}{\sqrt{m}}\right)
    \text.
\end{equation}
Above $I_1$ denotes the Bessel function of order $1$ and $1/m$ is the characteristic length scale of the Yukawa potential. 

With this modification, the holomorphic gradient flow may be used to compute the expectation value of $\theta(x)$ under the action Eq.~(\ref{eq:action}). The left panel of Figure~\ref{fig:yukawa} shows this expectation value---the probability of the particle being found to the right of the potential---as a function of time. The two different flow times shown are in good agreement with each other and with the result obtained from Trotterized time evolution. The transmission probability may be extracted as $P = \lim_{t \rightarrow \infty} \langle \theta(x)\rangle$.

The right-hand panel of Figure~\ref{fig:yukawa} shows the average phase obtained on flowed contours as a function of various flow times, for four different amounts of real-time evolution. In all Monte Carlo calculations $\delta = 0.1$ is used. The exact curve is computed by the method described at the top of this section with a Trotterized time step of $0.01$. 

As is often the case when the holomorphic gradient flow is used~\cite{Alexandru:2017oyw}, we find that the autocorrelation time of the Metropolis chain increases rapidly as a function of flow time. For flow times much beyond $T \gtrsim 0.4$, the autocorrelation time becomes such that even with $10^4$ samples reliable error bars cannot be obtained.

\section{Discussion}\label{sec:discussion}

We have shown how contour deformations methods---ordinarily limited to theories with a holomorphic Boltzmann factor---may be applied to singular actions. The core of the approach is that although the original action $S_{\mathrm{orig}}(z)$ is not holomorphic on $\mathbb C$, its restriction to the real line may in turn be extended to an action $S(z)$ defined on a different complex surface, which is holomorphic. Contour deformation methods may then be used with few modifications on $S(z)$.

From Figure~\ref{fig:yukawa} it is apparent that the contours constructed in this paper will not be sufficient to perform scattering calculations in more than one dimension. We leave the search for more effective contours to future work.

Contour deformation methods are not the only approach to the sign problem that depends on the holomorphicity of the integrand: complex Langevin makes this assumption as well~\cite{Aarts:2009uq}. It may be that a similar trick of defining a holomorphic integrand on a different complex surface will allow complex Langevin methods to be applied to such problems. Again we leave this question to future work.

Finally, despite their historical role in the development of contour deformation methods for the sign problem, we have not addressed the Lefschetz thimbles, nor saddle-point approximations, in this paper. It may be profitable to further develop the theory of saddle-point approximations and Lefschetz thimble regularizations in the context of non-holomorphic actions.

\begin{acknowledgments}
The view of the branch structure of $|x|$ as a degenerate case that of $\sqrt{(x-a)(x+a)}$ was pointed out by Tanmoy Bhattacharya.

S.L.~is supported by a Richard P.~Feynman fellowship from the LANL LDRD program. S.V.~is supported by the U.S. Department of Energy, Nuclear Physics Quantum Horizons program through the Early Career Award DE-SC0021892. Y.Y.~is supported by the INT's U.S. Department of Energy grant No.~DE-FG02-00ER41132.

Los Alamos National Laboratory is operated by Triad National Security, LLC, for the National Nuclear Security Administration of U.S. Department of Energy (Contract No.~89233218CNA000001).
\end{acknowledgments}

\bibliographystyle{apsrev4-2}
\bibliography{refs}

\begin{thebibliography}{26}%
\makeatletter
\providecommand \@ifxundefined [1]{%
 \@ifx{#1\undefined}
}%
\providecommand \@ifnum [1]{%
 \ifnum #1\expandafter \@firstoftwo
 \else \expandafter \@secondoftwo
 \fi
}%
\providecommand \@ifx [1]{%
 \ifx #1\expandafter \@firstoftwo
 \else \expandafter \@secondoftwo
 \fi
}%
\providecommand \natexlab [1]{#1}%
\providecommand \enquote  [1]{``#1''}%
\providecommand \bibnamefont  [1]{#1}%
\providecommand \bibfnamefont [1]{#1}%
\providecommand \citenamefont [1]{#1}%
\providecommand \href@noop [0]{\@secondoftwo}%
\providecommand \href [0]{\begingroup \@sanitize@url \@href}%
\providecommand \@href[1]{\@@startlink{#1}\@@href}%
\providecommand \@@href[1]{\endgroup#1\@@endlink}%
\providecommand \@sanitize@url [0]{\catcode `\\12\catcode `\$12\catcode `\&12\catcode `\#12\catcode `\^12\catcode `\_12\catcode `\%12\relax}%
\providecommand \@@startlink[1]{}%
\providecommand \@@endlink[0]{}%
\providecommand \url  [0]{\begingroup\@sanitize@url \@url }%
\providecommand \@url [1]{\endgroup\@href {#1}{\urlprefix }}%
\providecommand \urlprefix  [0]{URL }%
\providecommand \Eprint [0]{\href }%
\providecommand \doibase [0]{https://doi.org/}%
\providecommand \selectlanguage [0]{\@gobble}%
\providecommand \bibinfo  [0]{\@secondoftwo}%
\providecommand \bibfield  [0]{\@secondoftwo}%
\providecommand \translation [1]{[#1]}%
\providecommand \BibitemOpen [0]{}%
\providecommand \bibitemStop [0]{}%
\providecommand \bibitemNoStop [0]{.\EOS\space}%
\providecommand \EOS [0]{\spacefactor3000\relax}%
\providecommand \BibitemShut  [1]{\csname bibitem#1\endcsname}%
\let\auto@bib@innerbib\@empty
%</preamble>
\bibitem [{\citenamefont {Aarts}\ and\ \citenamefont {Stamatescu}(2008)}]{Aarts:2008rr}%
  \BibitemOpen
  \bibfield  {author} {\bibinfo {author} {\bibfnamefont {G.}~\bibnamefont {Aarts}}\ and\ \bibinfo {author} {\bibfnamefont {I.-O.}\ \bibnamefont {Stamatescu}},\ }\href {https://doi.org/10.1088/1126-6708/2008/09/018} {\bibfield  {journal} {\bibinfo  {journal} {JHEP}\ }\textbf {\bibinfo {volume} {09}},\ \bibinfo {pages} {018}},\ \Eprint {https://arxiv.org/abs/0807.1597} {arXiv:0807.1597 [hep-lat]} \BibitemShut {NoStop}%
\bibitem [{\citenamefont {Langfeld}\ and\ \citenamefont {Lucini}(2016)}]{Langfeld:2016mct}%
  \BibitemOpen
  \bibfield  {author} {\bibinfo {author} {\bibfnamefont {K.}~\bibnamefont {Langfeld}}\ and\ \bibinfo {author} {\bibfnamefont {B.}~\bibnamefont {Lucini}},\ }\bibfield  {booktitle} {\emph {\bibinfo {booktitle} {{Proceedings, International Meeting Excited QCD 2016: Costa da Caparica, Portugal, March 6-12, 2016}}},\ }\href {https://doi.org/10.5506/APhysPolBSupp.9.503} {\bibfield  {journal} {\bibinfo  {journal} {Acta Phys. Polon. Supp.}\ }\textbf {\bibinfo {volume} {9}},\ \bibinfo {pages} {503} (\bibinfo {year} {2016})},\ \Eprint {https://arxiv.org/abs/1606.03879} {arXiv:1606.03879 [hep-lat]} \BibitemShut {NoStop}%
%%CITATION = ARXIV:1606.03879;%%
\bibitem [{\citenamefont {Alexandru}\ \emph {et~al.}(2005)\citenamefont {Alexandru}, \citenamefont {Faber}, \citenamefont {Horvath},\ and\ \citenamefont {Liu}}]{Alexandru:2005ix}%
  \BibitemOpen
  \bibfield  {author} {\bibinfo {author} {\bibfnamefont {A.}~\bibnamefont {Alexandru}}, \bibinfo {author} {\bibfnamefont {M.}~\bibnamefont {Faber}}, \bibinfo {author} {\bibfnamefont {I.}~\bibnamefont {Horvath}},\ and\ \bibinfo {author} {\bibfnamefont {K.-F.}\ \bibnamefont {Liu}},\ }\href {https://doi.org/10.1103/PhysRevD.72.114513} {\bibfield  {journal} {\bibinfo  {journal} {Phys. Rev.}\ }\textbf {\bibinfo {volume} {D72}},\ \bibinfo {pages} {114513} (\bibinfo {year} {2005})},\ \Eprint {https://arxiv.org/abs/hep-lat/0507020} {arXiv:hep-lat/0507020 [hep-lat]} \BibitemShut {NoStop}%
%%CITATION = HEP-LAT/0507020;%%
\bibitem [{\citenamefont {de~Forcrand}\ and\ \citenamefont {Kratochvila}(2006)}]{deForcrand:2006ec}%
  \BibitemOpen
  \bibfield  {author} {\bibinfo {author} {\bibfnamefont {P.}~\bibnamefont {de~Forcrand}}\ and\ \bibinfo {author} {\bibfnamefont {S.}~\bibnamefont {Kratochvila}},\ }\bibfield  {booktitle} {\emph {\bibinfo {booktitle} {{Hadron physics, proceedings of the Workshop on Computational Hadron Physics, University of Cyprus, Nicosia, Cyprus, 14-17 September 2005}}},\ }\href {https://doi.org/10.1016/j.nuclphysbps.2006.01.007} {\bibfield  {journal} {\bibinfo  {journal} {Nucl. Phys. Proc. Suppl.}\ }\textbf {\bibinfo {volume} {153}},\ \bibinfo {pages} {62} (\bibinfo {year} {2006})},\ \bibinfo {note} {[,62(2006)]},\ \Eprint {https://arxiv.org/abs/hep-lat/0602024} {arXiv:hep-lat/0602024 [hep-lat]} \BibitemShut {NoStop}%
%%CITATION = HEP-LAT/0602024;%%
\bibitem [{\citenamefont {Fodor}\ and\ \citenamefont {Katz}(2002)}]{Fodor:2001au}%
  \BibitemOpen
  \bibfield  {author} {\bibinfo {author} {\bibfnamefont {Z.}~\bibnamefont {Fodor}}\ and\ \bibinfo {author} {\bibfnamefont {S.~D.}\ \bibnamefont {Katz}},\ }\href {https://doi.org/10.1016/S0370-2693(02)01583-6} {\bibfield  {journal} {\bibinfo  {journal} {Phys. Lett.}\ }\textbf {\bibinfo {volume} {B534}},\ \bibinfo {pages} {87} (\bibinfo {year} {2002})},\ \Eprint {https://arxiv.org/abs/hep-lat/0104001} {arXiv:hep-lat/0104001 [hep-lat]} \BibitemShut {NoStop}%
%%CITATION = HEP-LAT/0104001;%%
\bibitem [{\citenamefont {Allton}\ \emph {et~al.}(2002)\citenamefont {Allton}, \citenamefont {Ejiri}, \citenamefont {Hands}, \citenamefont {Kaczmarek}, \citenamefont {Karsch}, \citenamefont {Laermann}, \citenamefont {Schmidt},\ and\ \citenamefont {Scorzato}}]{Allton:2002zi}%
  \BibitemOpen
  \bibfield  {author} {\bibinfo {author} {\bibfnamefont {C.~R.}\ \bibnamefont {Allton}}, \bibinfo {author} {\bibfnamefont {S.}~\bibnamefont {Ejiri}}, \bibinfo {author} {\bibfnamefont {S.~J.}\ \bibnamefont {Hands}}, \bibinfo {author} {\bibfnamefont {O.}~\bibnamefont {Kaczmarek}}, \bibinfo {author} {\bibfnamefont {F.}~\bibnamefont {Karsch}}, \bibinfo {author} {\bibfnamefont {E.}~\bibnamefont {Laermann}}, \bibinfo {author} {\bibfnamefont {C.}~\bibnamefont {Schmidt}},\ and\ \bibinfo {author} {\bibfnamefont {L.}~\bibnamefont {Scorzato}},\ }\href {https://doi.org/10.1103/PhysRevD.66.074507} {\bibfield  {journal} {\bibinfo  {journal} {Phys. Rev.}\ }\textbf {\bibinfo {volume} {D66}},\ \bibinfo {pages} {074507} (\bibinfo {year} {2002})},\ \Eprint {https://arxiv.org/abs/hep-lat/0204010} {arXiv:hep-lat/0204010 [hep-lat]} \BibitemShut {NoStop}%
%%CITATION = HEP-LAT/0204010;%%
\bibitem [{\citenamefont {Chandrasekharan}(2013)}]{Chandrasekharan:2013rpa}%
  \BibitemOpen
  \bibfield  {author} {\bibinfo {author} {\bibfnamefont {S.}~\bibnamefont {Chandrasekharan}},\ }\href {https://doi.org/10.1140/epja/i2013-13090-y} {\bibfield  {journal} {\bibinfo  {journal} {Eur. Phys. J.}\ }\textbf {\bibinfo {volume} {A49}},\ \bibinfo {pages} {90} (\bibinfo {year} {2013})},\ \Eprint {https://arxiv.org/abs/1304.4900} {arXiv:1304.4900 [hep-lat]} \BibitemShut {NoStop}%
%%CITATION = ARXIV:1304.4900;%%
\bibitem [{\citenamefont {de~Forcrand}\ and\ \citenamefont {Philipsen}(2007)}]{deForcrand:2006pv}%
  \BibitemOpen
  \bibfield  {author} {\bibinfo {author} {\bibfnamefont {P.}~\bibnamefont {de~Forcrand}}\ and\ \bibinfo {author} {\bibfnamefont {O.}~\bibnamefont {Philipsen}},\ }\href {https://doi.org/10.1088/1126-6708/2007/01/077} {\bibfield  {journal} {\bibinfo  {journal} {JHEP}\ }\textbf {\bibinfo {volume} {01}},\ \bibinfo {pages} {077}},\ \Eprint {https://arxiv.org/abs/hep-lat/0607017} {arXiv:hep-lat/0607017 [hep-lat]} \BibitemShut {NoStop}%
%%CITATION = HEP-LAT/0607017;%%
\bibitem [{\citenamefont {Alexandru}\ \emph {et~al.}(2022)\citenamefont {Alexandru}, \citenamefont {Basar}, \citenamefont {Bedaque},\ and\ \citenamefont {Warrington}}]{Alexandru:2020wrj}%
  \BibitemOpen
  \bibfield  {author} {\bibinfo {author} {\bibfnamefont {A.}~\bibnamefont {Alexandru}}, \bibinfo {author} {\bibfnamefont {G.}~\bibnamefont {Basar}}, \bibinfo {author} {\bibfnamefont {P.~F.}\ \bibnamefont {Bedaque}},\ and\ \bibinfo {author} {\bibfnamefont {N.~C.}\ \bibnamefont {Warrington}},\ }\href {https://doi.org/10.1103/RevModPhys.94.015006} {\bibfield  {journal} {\bibinfo  {journal} {Rev. Mod. Phys.}\ }\textbf {\bibinfo {volume} {94}},\ \bibinfo {pages} {015006} (\bibinfo {year} {2022})},\ \Eprint {https://arxiv.org/abs/2007.05436} {arXiv:2007.05436 [hep-lat]} \BibitemShut {NoStop}%
\bibitem [{\citenamefont {Lawrence}\ \emph {et~al.}(2022)\citenamefont {Lawrence}, \citenamefont {Oh},\ and\ \citenamefont {Yamauchi}}]{Lawrence:2022afv}%
  \BibitemOpen
  \bibfield  {author} {\bibinfo {author} {\bibfnamefont {S.}~\bibnamefont {Lawrence}}, \bibinfo {author} {\bibfnamefont {H.}~\bibnamefont {Oh}},\ and\ \bibinfo {author} {\bibfnamefont {Y.}~\bibnamefont {Yamauchi}},\ }\href {https://doi.org/10.1103/PhysRevD.106.114503} {\bibfield  {journal} {\bibinfo  {journal} {Phys. Rev. D}\ }\textbf {\bibinfo {volume} {106}},\ \bibinfo {pages} {114503} (\bibinfo {year} {2022})},\ \Eprint {https://arxiv.org/abs/2205.12303} {arXiv:2205.12303 [hep-lat]} \BibitemShut {NoStop}%
\bibitem [{\citenamefont {Cristoforetti}\ \emph {et~al.}(2012)\citenamefont {Cristoforetti}, \citenamefont {Di~Renzo},\ and\ \citenamefont {Scorzato}}]{Cristoforetti:2012su}%
  \BibitemOpen
  \bibfield  {author} {\bibinfo {author} {\bibfnamefont {M.}~\bibnamefont {Cristoforetti}}, \bibinfo {author} {\bibfnamefont {F.}~\bibnamefont {Di~Renzo}},\ and\ \bibinfo {author} {\bibfnamefont {L.}~\bibnamefont {Scorzato}} (\bibinfo {collaboration} {AuroraScience}),\ }\href {https://doi.org/10.1103/PhysRevD.86.074506} {\bibfield  {journal} {\bibinfo  {journal} {Phys. Rev. D}\ }\textbf {\bibinfo {volume} {86}},\ \bibinfo {pages} {074506} (\bibinfo {year} {2012})},\ \Eprint {https://arxiv.org/abs/1205.3996} {arXiv:1205.3996 [hep-lat]} \BibitemShut {NoStop}%
\bibitem [{\citenamefont {Alexandru}\ \emph {et~al.}(2017{\natexlab{a}})\citenamefont {Alexandru}, \citenamefont {Bedaque}, \citenamefont {Lamm},\ and\ \citenamefont {Lawrence}}]{Alexandru:2017czx}%
  \BibitemOpen
  \bibfield  {author} {\bibinfo {author} {\bibfnamefont {A.}~\bibnamefont {Alexandru}}, \bibinfo {author} {\bibfnamefont {P.~F.}\ \bibnamefont {Bedaque}}, \bibinfo {author} {\bibfnamefont {H.}~\bibnamefont {Lamm}},\ and\ \bibinfo {author} {\bibfnamefont {S.}~\bibnamefont {Lawrence}},\ }\href {https://doi.org/10.1103/PhysRevD.96.094505} {\bibfield  {journal} {\bibinfo  {journal} {Phys. Rev.}\ }\textbf {\bibinfo {volume} {D96}},\ \bibinfo {pages} {094505} (\bibinfo {year} {2017}{\natexlab{a}})},\ \Eprint {https://arxiv.org/abs/1709.01971} {arXiv:1709.01971 [hep-lat]} \BibitemShut {NoStop}%
%%CITATION = ARXIV:1709.01971;%%
\bibitem [{\citenamefont {Alexandru}\ \emph {et~al.}(2016{\natexlab{a}})\citenamefont {Alexandru}, \citenamefont {Basar},\ and\ \citenamefont {Bedaque}}]{Alexandru:2015xva}%
  \BibitemOpen
  \bibfield  {author} {\bibinfo {author} {\bibfnamefont {A.}~\bibnamefont {Alexandru}}, \bibinfo {author} {\bibfnamefont {G.}~\bibnamefont {Basar}},\ and\ \bibinfo {author} {\bibfnamefont {P.}~\bibnamefont {Bedaque}},\ }\href {https://doi.org/10.1103/PhysRevD.93.014504} {\bibfield  {journal} {\bibinfo  {journal} {Phys. Rev. D}\ }\textbf {\bibinfo {volume} {93}},\ \bibinfo {pages} {014504} (\bibinfo {year} {2016}{\natexlab{a}})},\ \Eprint {https://arxiv.org/abs/1510.03258} {arXiv:1510.03258 [hep-lat]} \BibitemShut {NoStop}%
\bibitem [{\citenamefont {Lawrence}\ and\ \citenamefont {Yamauchi}(2021)}]{Lawrence:2021izu}%
  \BibitemOpen
  \bibfield  {author} {\bibinfo {author} {\bibfnamefont {S.}~\bibnamefont {Lawrence}}\ and\ \bibinfo {author} {\bibfnamefont {Y.}~\bibnamefont {Yamauchi}},\ }\href {https://doi.org/10.1103/PhysRevD.103.114509} {\bibfield  {journal} {\bibinfo  {journal} {Phys. Rev. D}\ }\textbf {\bibinfo {volume} {103}},\ \bibinfo {pages} {114509} (\bibinfo {year} {2021})},\ \Eprint {https://arxiv.org/abs/2101.05755} {arXiv:2101.05755 [hep-lat]} \BibitemShut {NoStop}%
\bibitem [{\citenamefont {Alexandru}\ \emph {et~al.}(2018{\natexlab{a}})\citenamefont {Alexandru}, \citenamefont {Bedaque}, \citenamefont {Lamm}, \citenamefont {Lawrence},\ and\ \citenamefont {Warrington}}]{Alexandru:2018ddf}%
  \BibitemOpen
  \bibfield  {author} {\bibinfo {author} {\bibfnamefont {A.}~\bibnamefont {Alexandru}}, \bibinfo {author} {\bibfnamefont {P.~F.}\ \bibnamefont {Bedaque}}, \bibinfo {author} {\bibfnamefont {H.}~\bibnamefont {Lamm}}, \bibinfo {author} {\bibfnamefont {S.}~\bibnamefont {Lawrence}},\ and\ \bibinfo {author} {\bibfnamefont {N.~C.}\ \bibnamefont {Warrington}},\ }\href {https://doi.org/10.1103/PhysRevLett.121.191602} {\bibfield  {journal} {\bibinfo  {journal} {Phys. Rev. Lett.}\ }\textbf {\bibinfo {volume} {121}},\ \bibinfo {pages} {191602} (\bibinfo {year} {2018}{\natexlab{a}})},\ \Eprint {https://arxiv.org/abs/1808.09799} {arXiv:1808.09799 [hep-lat]} \BibitemShut {NoStop}%
%%CITATION = ARXIV:1808.09799;%%
\bibitem [{\citenamefont {Alexandru}\ \emph {et~al.}(2018{\natexlab{b}})\citenamefont {Alexandru}, \citenamefont {Bedaque}, \citenamefont {Lamm},\ and\ \citenamefont {Lawrence}}]{Alexandru:2018fqp}%
  \BibitemOpen
  \bibfield  {author} {\bibinfo {author} {\bibfnamefont {A.}~\bibnamefont {Alexandru}}, \bibinfo {author} {\bibfnamefont {P.~F.}\ \bibnamefont {Bedaque}}, \bibinfo {author} {\bibfnamefont {H.}~\bibnamefont {Lamm}},\ and\ \bibinfo {author} {\bibfnamefont {S.}~\bibnamefont {Lawrence}},\ }\href {https://doi.org/10.1103/PhysRevD.97.094510} {\bibfield  {journal} {\bibinfo  {journal} {Phys. Rev. D}\ }\textbf {\bibinfo {volume} {97}},\ \bibinfo {pages} {094510} (\bibinfo {year} {2018}{\natexlab{b}})},\ \Eprint {https://arxiv.org/abs/1804.00697} {arXiv:1804.00697 [hep-lat]} \BibitemShut {NoStop}%
\bibitem [{\citenamefont {Alexandru}\ \emph {et~al.}(2018{\natexlab{c}})\citenamefont {Alexandru}, \citenamefont {Ba\c{s}ar}, \citenamefont {Bedaque}, \citenamefont {Lamm},\ and\ \citenamefont {Lawrence}}]{Alexandru:2018ngw}%
  \BibitemOpen
  \bibfield  {author} {\bibinfo {author} {\bibfnamefont {A.}~\bibnamefont {Alexandru}}, \bibinfo {author} {\bibfnamefont {G.}~\bibnamefont {Ba\c{s}ar}}, \bibinfo {author} {\bibfnamefont {P.~F.}\ \bibnamefont {Bedaque}}, \bibinfo {author} {\bibfnamefont {H.}~\bibnamefont {Lamm}},\ and\ \bibinfo {author} {\bibfnamefont {S.}~\bibnamefont {Lawrence}},\ }\href {https://doi.org/10.1103/PhysRevD.98.034506} {\bibfield  {journal} {\bibinfo  {journal} {Phys. Rev. D}\ }\textbf {\bibinfo {volume} {98}},\ \bibinfo {pages} {034506} (\bibinfo {year} {2018}{\natexlab{c}})},\ \Eprint {https://arxiv.org/abs/1807.02027} {arXiv:1807.02027 [hep-lat]} \BibitemShut {NoStop}%
\bibitem [{\citenamefont {Mori}\ \emph {et~al.}(2017)\citenamefont {Mori}, \citenamefont {Kashiwa},\ and\ \citenamefont {Ohnishi}}]{Mori:2017pne}%
  \BibitemOpen
  \bibfield  {author} {\bibinfo {author} {\bibfnamefont {Y.}~\bibnamefont {Mori}}, \bibinfo {author} {\bibfnamefont {K.}~\bibnamefont {Kashiwa}},\ and\ \bibinfo {author} {\bibfnamefont {A.}~\bibnamefont {Ohnishi}},\ }\href {https://doi.org/10.1103/PhysRevD.96.111501} {\bibfield  {journal} {\bibinfo  {journal} {Phys. Rev. D}\ }\textbf {\bibinfo {volume} {96}},\ \bibinfo {pages} {111501} (\bibinfo {year} {2017})},\ \Eprint {https://arxiv.org/abs/1705.05605} {arXiv:1705.05605 [hep-lat]} \BibitemShut {NoStop}%
\bibitem [{\citenamefont {Kashiwa}\ and\ \citenamefont {Mori}(2020)}]{Kashiwa:2020brj}%
  \BibitemOpen
  \bibfield  {author} {\bibinfo {author} {\bibfnamefont {K.}~\bibnamefont {Kashiwa}}\ and\ \bibinfo {author} {\bibfnamefont {Y.}~\bibnamefont {Mori}},\ }\href {https://doi.org/10.1103/PhysRevD.102.054519} {\bibfield  {journal} {\bibinfo  {journal} {Phys. Rev. D}\ }\textbf {\bibinfo {volume} {102}},\ \bibinfo {pages} {054519} (\bibinfo {year} {2020})},\ \Eprint {https://arxiv.org/abs/2007.04167} {arXiv:2007.04167 [hep-lat]} \BibitemShut {NoStop}%
\bibitem [{\citenamefont {Kashiwa}\ \emph {et~al.}(2023)\citenamefont {Kashiwa}, \citenamefont {Namekawa}, \citenamefont {Ohnishi},\ and\ \citenamefont {Takase}}]{Kashiwa:2023dfx}%
  \BibitemOpen
  \bibfield  {author} {\bibinfo {author} {\bibfnamefont {K.}~\bibnamefont {Kashiwa}}, \bibinfo {author} {\bibfnamefont {Y.}~\bibnamefont {Namekawa}}, \bibinfo {author} {\bibfnamefont {A.}~\bibnamefont {Ohnishi}},\ and\ \bibinfo {author} {\bibfnamefont {H.}~\bibnamefont {Takase}},\ }\href@noop {} {\bibinfo {title} {{Application of the path optimization method to a discrete spin system}}} (\bibinfo {year} {2023}),\ \Eprint {https://arxiv.org/abs/2309.06018} {arXiv:2309.06018 [hep-lat]} \BibitemShut {NoStop}%
\bibitem [{\citenamefont {Kanwar}\ \emph {et~al.}(2023)\citenamefont {Kanwar}, \citenamefont {Lovato}, \citenamefont {Rocco},\ and\ \citenamefont {Wagman}}]{Kanwar:2023otc}%
  \BibitemOpen
  \bibfield  {author} {\bibinfo {author} {\bibfnamefont {G.}~\bibnamefont {Kanwar}}, \bibinfo {author} {\bibfnamefont {A.}~\bibnamefont {Lovato}}, \bibinfo {author} {\bibfnamefont {N.}~\bibnamefont {Rocco}},\ and\ \bibinfo {author} {\bibfnamefont {M.}~\bibnamefont {Wagman}},\ }\href@noop {} {\bibinfo {title} {{Mitigating Green's function Monte Carlo signal-to-noise problems using contour deformations}}} (\bibinfo {year} {2023}),\ \Eprint {https://arxiv.org/abs/2304.03229} {arXiv:2304.03229 [nucl-th]} \BibitemShut {NoStop}%
\bibitem [{\citenamefont {Alexandru}\ \emph {et~al.}(2016{\natexlab{b}})\citenamefont {Alexandru}, \citenamefont {Basar}, \citenamefont {Bedaque}, \citenamefont {Ridgway},\ and\ \citenamefont {Warrington}}]{Alexandru:2015sua}%
  \BibitemOpen
  \bibfield  {author} {\bibinfo {author} {\bibfnamefont {A.}~\bibnamefont {Alexandru}}, \bibinfo {author} {\bibfnamefont {G.}~\bibnamefont {Basar}}, \bibinfo {author} {\bibfnamefont {P.~F.}\ \bibnamefont {Bedaque}}, \bibinfo {author} {\bibfnamefont {G.~W.}\ \bibnamefont {Ridgway}},\ and\ \bibinfo {author} {\bibfnamefont {N.~C.}\ \bibnamefont {Warrington}},\ }\href {https://doi.org/10.1007/JHEP05(2016)053} {\bibfield  {journal} {\bibinfo  {journal} {JHEP}\ }\textbf {\bibinfo {volume} {05}},\ \bibinfo {pages} {053}},\ \Eprint {https://arxiv.org/abs/1512.08764} {arXiv:1512.08764 [hep-lat]} \BibitemShut {NoStop}%
\bibitem [{\citenamefont {Alexandru}\ \emph {et~al.}(2016{\natexlab{c}})\citenamefont {Alexandru}, \citenamefont {Basar}, \citenamefont {Bedaque}, \citenamefont {Vartak},\ and\ \citenamefont {Warrington}}]{Alexandru:2016gsd}%
  \BibitemOpen
  \bibfield  {author} {\bibinfo {author} {\bibfnamefont {A.}~\bibnamefont {Alexandru}}, \bibinfo {author} {\bibfnamefont {G.}~\bibnamefont {Basar}}, \bibinfo {author} {\bibfnamefont {P.~F.}\ \bibnamefont {Bedaque}}, \bibinfo {author} {\bibfnamefont {S.}~\bibnamefont {Vartak}},\ and\ \bibinfo {author} {\bibfnamefont {N.~C.}\ \bibnamefont {Warrington}},\ }\href {https://doi.org/10.1103/PhysRevLett.117.081602} {\bibfield  {journal} {\bibinfo  {journal} {Phys. Rev. Lett.}\ }\textbf {\bibinfo {volume} {117}},\ \bibinfo {pages} {081602} (\bibinfo {year} {2016}{\natexlab{c}})},\ \Eprint {https://arxiv.org/abs/1605.08040} {arXiv:1605.08040 [hep-lat]} \BibitemShut {NoStop}%
\bibitem [{\citenamefont {Alexandru}\ \emph {et~al.}(2017{\natexlab{b}})\citenamefont {Alexandru}, \citenamefont {Basar}, \citenamefont {Bedaque},\ and\ \citenamefont {Ridgway}}]{Alexandru:2017lqr}%
  \BibitemOpen
  \bibfield  {author} {\bibinfo {author} {\bibfnamefont {A.}~\bibnamefont {Alexandru}}, \bibinfo {author} {\bibfnamefont {G.}~\bibnamefont {Basar}}, \bibinfo {author} {\bibfnamefont {P.~F.}\ \bibnamefont {Bedaque}},\ and\ \bibinfo {author} {\bibfnamefont {G.~W.}\ \bibnamefont {Ridgway}},\ }\href {https://doi.org/10.1103/PhysRevD.95.114501} {\bibfield  {journal} {\bibinfo  {journal} {Phys. Rev. D}\ }\textbf {\bibinfo {volume} {95}},\ \bibinfo {pages} {114501} (\bibinfo {year} {2017}{\natexlab{b}})},\ \Eprint {https://arxiv.org/abs/1704.06404} {arXiv:1704.06404 [hep-lat]} \BibitemShut {NoStop}%
\bibitem [{\citenamefont {Alexandru}\ \emph {et~al.}(2017{\natexlab{c}})\citenamefont {Alexandru}, \citenamefont {Basar}, \citenamefont {Bedaque},\ and\ \citenamefont {Warrington}}]{Alexandru:2017oyw}%
  \BibitemOpen
  \bibfield  {author} {\bibinfo {author} {\bibfnamefont {A.}~\bibnamefont {Alexandru}}, \bibinfo {author} {\bibfnamefont {G.}~\bibnamefont {Basar}}, \bibinfo {author} {\bibfnamefont {P.~F.}\ \bibnamefont {Bedaque}},\ and\ \bibinfo {author} {\bibfnamefont {N.~C.}\ \bibnamefont {Warrington}},\ }\href {https://doi.org/10.1103/PhysRevD.96.034513} {\bibfield  {journal} {\bibinfo  {journal} {Phys. Rev. D}\ }\textbf {\bibinfo {volume} {96}},\ \bibinfo {pages} {034513} (\bibinfo {year} {2017}{\natexlab{c}})},\ \Eprint {https://arxiv.org/abs/1703.02414} {arXiv:1703.02414 [hep-lat]} \BibitemShut {NoStop}%
\bibitem [{\citenamefont {Aarts}\ \emph {et~al.}(2010)\citenamefont {Aarts}, \citenamefont {Seiler},\ and\ \citenamefont {Stamatescu}}]{Aarts:2009uq}%
  \BibitemOpen
  \bibfield  {author} {\bibinfo {author} {\bibfnamefont {G.}~\bibnamefont {Aarts}}, \bibinfo {author} {\bibfnamefont {E.}~\bibnamefont {Seiler}},\ and\ \bibinfo {author} {\bibfnamefont {I.-O.}\ \bibnamefont {Stamatescu}},\ }\href {https://doi.org/10.1103/PhysRevD.81.054508} {\bibfield  {journal} {\bibinfo  {journal} {Phys. Rev. D}\ }\textbf {\bibinfo {volume} {81}},\ \bibinfo {pages} {054508} (\bibinfo {year} {2010})},\ \Eprint {https://arxiv.org/abs/0912.3360} {arXiv:0912.3360 [hep-lat]} \BibitemShut {NoStop}%
\end{thebibliography}%

\end{document}